\documentclass[twocolumn,showpacs,preprintnumbers,amsmath,amssymb,pre]{revtex4}

\usepackage{times,graphicx,amssymb,amsmath}

\begin{document}

\title{Emergence and resilience of social networks: a general theoretical framework}

\author{George C.M.A. Ehrhardt}
\email{gehrhard@ictp.trieste.it}
\author{Matteo Marsili}
\affiliation{The Abdus Salam ICTP, Strada Costiera 11, I-34014, Trieste. Italy.}
\author{Fernando Vega-Redondo}
\affiliation{Universidad de Alicante, Facultad de Economicas, Universidad de Alicante, 03071, Alicante. Spain.}
\altaffiliation[also ]{University of Essex, Wivenhoe Park, Colchester, CO4 3SQ, UK.}

\begin{abstract}
We introduce and study a general model of social network formation and evolution based on the concept of preferential link formation between similar nodes and increased similarity between connected nodes.  The model is studied numerically and analytically for three definitions of similarity.  In common with real-world social networks, we find coexistence of high and low connectivity phases and history dependence.  We suggest that the positive feedback between linking and similarity which is responsible for the model's behaviour is also an important mechanism in real social networks.
\end{abstract}

\maketitle

\section{Introduction}

There is a growing consensus among social scientists that many social
phenomena display an inherent network dimension. Not only are they \textquotedblleft embedded\textquotedblright\ in the underlying
social network \cite{Granov} but, reciprocally, the social network itself is
largely shaped by the evolution of those phenomena. The range of social
problems subject to these considerations is wide and important. It includes,
for example, the spread of crime \cite{Glaeser et al, Haynie} and other
social problems (e.g. teenage pregnancy \cite{Crane, Harding}), the rise of
industrial districts \cite{OECD, Saxenian, Granovetter et al}, and the
establishment of research collaborations, both scientific \cite{Newman,
Goyal et al} and industrial \cite{Hagedoorn,Kogut}. Throughout these cases,
there are a number of interesting observations worth highlighting:\medskip

(a) \textbf{Sharp transitions}:\emph{\ The shift from a sparse to a highly
connected network often unfolds rather \textquotedblleft
abruptly,\textquotedblright\ i.e. in a short timespan}. For example,
concerning the escalation of social pathologies in some neighborhoods of
large cities, Crane \cite{Crane} writes that \textquotedblleft ...if the
incidence [of the problem] reaches a critical point, the process of spread
will explode.\textquotedblright \ Also, considering the growth of 
research collaboration networks, Goyal \emph{et al.} \cite{Goyal et al}
report a steep increase in the per capita number of collaborations among
academic economists in the last three decades, while Hagerdoorn \cite{Hagedoorn}
reports an even sharper (ten-fold) increase for R\&D partnerships among
firms during the decade 1975-1985.\medskip

(b)\textbf{\ Resilience}:\emph{\ Once the transition to a highly connected
network has taken place, the network is robust, surviving even
a reversion to \textquotedblleft unfavorable\textquotedblright\ conditions}.
The case of California's Silicon
Valley, discussed in a classic account by Saxenian \cite{Saxenian},
illustrates this point well. Its thriving performance, even in the face of
the general crisis undergone by the computer industry in the 80's, 
has been largely attributed to the dense and flexible networks of collaboration
across individual actors that characterized it. 
Another intrinsically network-based example is the
rapid recent development of Open-Source software (e.g. Linux), a phenomenon
sustained against large odds by a dense web of collaboration and trust \cite
{Benkler}. Finally, as an example where \textquotedblleft
robustness\textquotedblright\ has negative rather than positive
implications,\ Crane \cite{Crane} describes the difficulty, even with vigorous social measures,
of improving a local neighborhood once crime and other
social pathologies have taken hold. \medskip 

(c) \textbf{Equilibrium co-existence}: \emph{Under apparently similar
environmental conditions, social networks may be found both in a dense or sparse state}. 
Again, a good illustration is provided by the
dual experience of poor neighborhoods in large cities \cite{Crane}, where
neither poverty nor other socio-economic conditions (e.g. ethnic
composition) can alone explain  whether or not there is degradation into a
ghetto with rampant social problems. Returning to R\&D
partnerships, empirical evidence \cite{Hagedoorn} shows a very
polarized situation, almost all R\&D partnerships taking
place in a few (high-technology) industries.  Even within those
industries, partnerships are almost exclusively between a small
subset of firms in (highly advanced) countries.\footnote{
Specifically, Hagerdon \cite{Hagedoorn} reports that 99\% of the R\&D
partnerships worldwide are conducted among firms in the so-called Triad:
North America, Europe and Japan.} \bigskip

From a theoretical viewpoint, the above discussion raises the question of
whether there is some common mechanism at work in the dynamics of social
networks that, in a wide variety of different scenarios, produces the three
features explained above: (a) discontinuous phase transitions, (b)
resilience, and (c) equilibrium coexistence. Our aim in this paper is to
shed light on this question within a general framework that is flexible
enough to accommodate, under alternative concrete specifications, a rich
range of social-network dynamics.

The recent literature on complex networks has largely focused on
understanding what are the generic properties arising in networks under
different link formation mechanisms. Those properties are important to
gain a proper theoretical grasp of many network phenomena and also provide
useful guiding principles for empirical research.  The analysis, however, has been mostly static, largely concerned with features such as small-world  \cite{WattsStrogatz} or scale-free \cite{Barabasi} networks. 
In contrast, our approach in this paper to the issue of network formation is
intrinsically dynamic, the steady state being a balance of link formation and removal.

We consider a set of agents -- be they individuals or organizations -- who establish bilateral
interactions (links) when profitable. The network evolves under changing
conditions. That is, the favorable circumstances that led at some point to
the formation of a particular link may later on deteriorate, causing
that link's removal.  Hence volatility (exogenous or endogenous) is a key disruptive element in the
dynamics.  Concurrently, new opportunities arise that favour
the formation of new links.  Whether linking occurs depends on factors related 
to the similarity or proximity of
the two parties. For example, in cases where trust is essential in
the establishment of new relationships (e.g. in crime or trade networks),
linking may be facilitated by common acquaintances or by
the existence of a chain of acquaintances joining the two parties. In other
cases (e.g. in R\&D or scientific networks), a common
language, methodology, or comparable level of technical competence may be required for
the link to be feasible or fruitful to both parties.

In a nutshell, our model conceives the dynamics of the network as a
struggle between volatility (that causes link decay) on the one hand, and
the creation of new links (that is dependent on similarity) on the other.
The model must also specify the dynamics governing
inter-node similarity.  A reasonable assumption in this respect is
that such similarity is enhanced by close interaction, as reflected by the
social network. For example, a firm (or researcher) benefits from 
collaborating with a similarly advanced partner, or individuals who interact regularly tend to
converge on their social norms and other standards of behavior.

We study different specifications of the general framework, each one
embodying alternative forms of the intuitive idea that \textquotedblleft
interaction promotes similarity.\textquotedblright\ Our main finding is that
in all of these different cases the network dynamics exhibits, over a wide
range of parameters, the type of phenomenology discussed above. The
essential mechanism at work is a positive feedback between link creation and
internode similarity, these two factors each exerting a
positive effect on the other. Feedback forces of this kind appear to operate
in the dynamics of many social networks. We show that they are sufficient to
produce the sharp transitions, resilience, and equilibrium co-existence
that, as explained, are salient features of many social phenomena.

\section{The model}

Consider a set ${\cal N }=\{1,\ldots ,n \}$ of agents, whose interactions evolve in
continuous time $t$.  Their network of interaction at some $t$ is
described by a non-directed graph $g(t)\subset \{ij:i\in {\cal N },~j\in {\cal N }\}$, where 
$ij(\equiv ji)\in g(t)$ iff a link exists between agents $i$
and $j$. The network evolves in the following manner. Firstly, each node $i$
receives an opportunity to form a link with a node $j$, randomly drawn from 
${\cal N }$ ($i \ne j$), at rate $1$ (i.e. with a probability $dt$ in a time interval $[t,t+dt)$
). If this link $ij$ is not already in place, it forms with probability

\begin{equation}
P\{ij\rightarrow g(t)\}=\left\{ 
\begin{array}{cc}
1 & \;\text{if }d_{ij}(t)\leq \bar{d} \\ 
\epsilon  & \;\text{if }d_{ij}(t)>\bar{d}
\end{array}
\right.   \label{Pform}
\end{equation}
where $d_{ij}(t)$ is the \textquotedblleft distance\textquotedblright\ (to be
specified later) between $i$ and $j$ prevailing at $t.$  Thus if $i$ and 
$j$ are close, in the sense that their distance is no higher than some given
threshold $\bar{d}$, the link forms at rate $1$; otherwise, it forms at a
much smaller rate $\epsilon \ll 1$. Secondly, each
existing link $ij\in g(t)$ decays at rate $\lambda$.  That is, each link
in the network disappears with probability $\lambda dt$ in a
time interval $[t,t+dt)$.

We shall discuss three different specifications of the distance 
$d_{ij}$, each capturing different aspects that may be relevant for
socio-economic interactions.  Consider first the simplest possible such
specification where $d_{ij}(t)$ is the (geodesic) distance
between $i$ and $j$ on the graph $g(t)$, neighbors $j$ of $i$ having $
d_{ij}(t)=1$, neighbors of the neighbors of $i$ (which are not neighbors of $
i$) having $d_{ij}(t)=2,$ and so on.  If no path joins $i$ and $j$ we set $
d_{ij}(t)=\infty $.

This specific model describes a situation where the formation of new links
is strongly influenced by proximity on the graph. It is a simple manifestation 
of our general idea that close interaction brings
about similarity -- here the two metrics coincide. When 
$\bar{d}>n-1$, the link formation process discriminates between
agents belonging to the same network component (which are joined by at least
one path of links in $g$) and agents in different components. Distinct
components of the graph may, for example, represent different social groups.
Then Eq. (\ref{Pform}) captures the fact that belonging to the same social
group is important in the creation of new links (say, because it facilitates
control or reciprocity \cite{Coleman,An}). 

\begin{figure}[tbp]
\includegraphics[width=0.9\columnwidth]{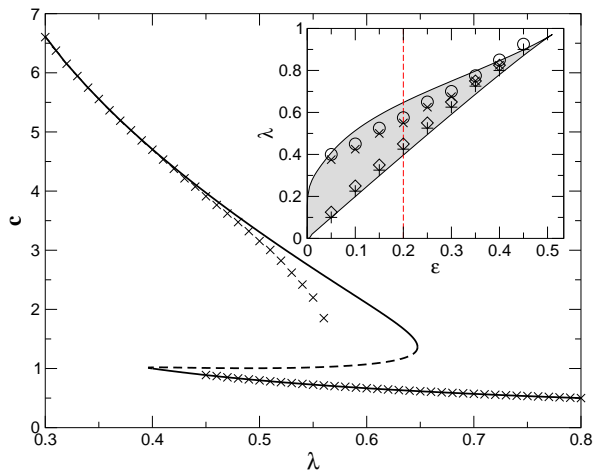}
\caption{
\label{fig_giant}
Mean degree $c$ as a function of $\lambda $ for $\epsilon =0.2$ 
when $d_{ij}$ is the distance on the graph and $\bar{d}>n-1$.
The results of a mean field theory for $n=\infty $ (solid line) is compared
to numerical simulations ($\times $) starting from both low and high connected states with 
$n=20000$. The dashed line
corresponds to an unstable solution of the mean field equations which
separates the basins of stability of the two solutions. Indeed the low
density state, for finite $n$, \textquotedblleft flips\textquotedblright\ to
the high density state when a random fluctuation in $c$ brings the system
across the stability boundary (i.e. when a sizable giant component forms).
These fluctuations become more and more rare as $n$ increases. \emph{Inset}
: Phase diagram in mean field theory. Coexistence occurs in the shaded
region whereas below (above) only the dense (sparse) network phase is
stable. Numerical simulations (symbols) agree qualitatively with the mean field
prediction. The high (low) density state is stable up (down) to the points
marked with $\times$ ($\diamond$) and is unstable at points marked with $\circ $ ($+$). 
The behavior of $c$ along the dashed line is reported in the
main figure.  
}
\end{figure}

Consider first what happens when $\lambda$ is large.  Let $c$ be the average 
connectivity (number of links per node) in the network.  The
average rate $n \lambda c/2$ of link removal is very high when $c$ is
significant. Consequently, we expect to have a very low $c$, which in turn
implies that the population should be fragmented into many small groups.
Under these circumstances, the likelihood that an agent $i$
\textquotedblleft meets\textquotedblright\ an agent $j$ in the same
component is negligible for large populations, and therefore new links are
created at a rate almost equal to $n \epsilon $. Invoking a simple
balance between link creation and link destruction, the average number of
neighbors of an agent is expected to be $c\simeq 2 \epsilon /\lambda$, as
is indeed found in our simulations (Fig \ref{fig_giant}).

As $\lambda $ decreases, the network density $c$ increases
gradually, but then, at a critical value $\lambda _{1},$ it makes a
discontinuous jump (Fig. \ref{fig_giant}) to a state containing a large and
densely interconnected community covering a finite fraction of the
population (the giant component).  Naturally, if volatility $\lambda $ decreases further, the
network becomes even more densely connected.  But, remarkably, if volatility
increases back again beyond the transition point $\lambda _{1},$ the dense
network remains stable.  The dense network dissolves back into a sparsely connected one 
only at a second point $\lambda _{2}$.  This
phenomenology characterizes a wide region of parameter space (see inset of
Fig. \ref{fig_giant}) and is qualitatively well reproduced by a simple mean
field approach (see appendix).

A similar phenomenology occurs when $\bar{d}=2$,
i.e. when links are preferentially formed with \textquotedblleft friends of
friends\textquotedblright , in an appropriate parameter range.\footnote{
Both $\epsilon $ and $\lambda $ must be comparable to the probability that
two arbitrary nodes $i$ and $j$ have $d_{ij}=2$, which is of order $1/n$ in
a network with finite degree.} This is reminiscent of a model that was
recently proposed \cite{PNAS} to describe a situation where (as e.g. in job
search \cite{Job}) agents find new linking opportunities through 
current partners. In \cite{PNAS} agents use their links to search
for new connections, whereas here existing links favor new link formation. 
In spite of this conceptual difference, the model in Ref. \cite
{PNAS} also features the phenomenology (a)-(c) above, i.e. sharp
transitions, resilience, and phase coexistence.

We now consider an alternative specialization of the general framework
where link formation requires some form of coordination,
synchronization, or compatibility.  For example, a profitable interaction may
fail to occur if the two parties do not agree on where and when to meet, or
if they do not speak the same languages, and/or adopt compatible
technologies and standards. In addition, it may well be that shared social
norms and codes enhance trust and thus are largely needed for fruitful
interaction.

To account for these considerations, we endow each agent with an
attribute $x_{i}$ which may take a finite number $q$ of different values, 
$x_{i}\in \{1,2,\ldots ,q\}$.  $x_{i}$ describes the internal state of
the agent, specifying e.g. its technological standard, language, or the
social norms she adopts.  The formation of a new link 
$ij$ requires that $i$ and $j$ display the same attribute, i.e. $x_{i}=x_{j}$. 
This is a particularization of the general Eq. (\ref{Pform}) with 
$d_{ij}=\delta_{x_{i},x_{j}}$ and $\bar{d}=0$.  For simplicity
we set $\epsilon =0$ since in the present
formulation there is always a finite probability that two nodes display the
same attribute and hence can link.  We assume each agent
revises its attribute at rate $\nu$, choosing $x_i$ dependent on its neighbours' $x_j$s according to:
\begin{equation}
P\{x_{i}(t)=x\}=\frac{1}{Z} \exp \left[ \beta \sum_{j:ij\in g(t)}\delta _{x,x_{j}(t)}
\right]   \label{potts}
\end{equation}
where $\beta $ tunes the tendency of agents to conform with 
their neighbors and $Z$ provides the normalisation.
This adjustment rule has a long tradition in physics \cite{Baxter} and also occurs
in the socio-economic literature as a model of coordination (or social conformity) 
under local interaction \cite{Blume,Durlauf,Young}. 
This is another manifestation of
our general idea that network-mediated contact favors internode similarity.
We focus on the case where such a similarity-enhancing dynamics proceeds at
a much faster rate than the network dynamics.  That is, $\nu \gg 1$ so that, 
at any given $t$ where the network $g(t)$ is about to change$,$ the attribute dynamics on the $x_{i}$
have relaxed to a stationary state.  The
statistics of this state is provided by the Potts model in
physics, which has been recently discussed for random graphs \cite{DGM,EM}.
We refer to the appendix for details and move
directly to discussing the results.

For a given $\beta $, under strong volatility ($\lambda \gg 1$) the
link density is very low, there is no giant component and agents $i$, $j$ chosen at random (for $n$ large) are not coordinated ($P( x_i=x_j ) =1/q$).  Hence links form at a node 
at rate $2/q$.  A simple balance of link formation and
decay rates implies that $c=2/(q\lambda )$ in this case.
When $\lambda $ decreases, network density $c$ increases.  First, it does so
gradually but at a critical point $\lambda _{1}(\beta )$ $c$ becomes sufficiently large 
that the $x_i$s within the giant component (whose existence is necessary for coordination) 
become coordinated.  Link formation increases since now 
$P( x_i=x_j ) > 1/q$ and this in turn increases the coordination.  This 
positive feedback causes a sharp transition to a coordinated, more highly connected state.
Once this sharp transition has
taken place, further decreases in $\lambda $ are simply reflected in gradual
increases in network density. On the other hand, subsequent changes of $
\lambda $ in the opposite direction are met by hysteresis.  That is, if $\lambda $ 
now grows starting at values below $\lambda _{1},$ the
network does \emph{not} revert to the sparse network at the latter
threshold. Rather, it remains in a dense state up to a larger value $\lambda
_{2}>\lambda _{1}$, sustained by the same positive feedback discussed above. 

This phenomenology, though induced by a different mechanism, is quite
similar in spirit to that reported in Fig. \ref{fig_giant} for the previous
model.  In the limit $\beta \rightarrow \infty $, the second model becomes equivalent
to the first one since with $\beta \rightarrow \infty $, all nodes in the
same component share the same value of $x_{i}(t)$, whilst the probability
to link two disconnected nodes is $\epsilon =1/q$. 
In fact, the roles of $1/\beta $ and $\lambda $ in the model are
analogous. If we fix $\lambda $ and parametrize the behavior of the model
through $1/\beta ,$ the same phenomena of discontinuous transitions,
hysteresis, and equilibrium co-existence occurs for corresponding threshold
values $1/\beta _{1}$ and $1/\beta _{2}$, analogous to $\lambda _{1}$ and $
\lambda _{2}$ in the former discussion.  

Finally, we consider a setup where $d_{ij}$ reflects proximity of nodes $i$, $j$ 
in terms of some continuous (non-negative) real attributes, 
$W_{i}(t)$, $W_{j}(t).$  These attributes could represent the level of
technical expertise of two firms involved in an R\&D partnership, or the
competence of two researchers involved in a joint project. It could also be
a measure of income or wealth that bears on the quality and prospects of a
bilateral relationship. Whatever the interpretation, it may be natural in
certain applications to posit that some process of diffusion tends to
equalize the levels displayed by neighboring agents. This idea is captured
by the following stochastic differential equation:
\begin{equation}
\frac{dW_{i}}{dt}=\nu \sum_{j:ij\in g}\left[ W_{j}(t)-W_{i}(t)\right]
+W_{i}(t)\eta _{i}(t)  \label{W}
\end{equation}
where $\eta _{i}(t)$ is uncorrelated white noise, i.e. $\langle {\eta
_{i}(t)\eta _{j}(t^{\prime })}\rangle =D\delta _{ij}\delta (t-t^{\prime })$.
The first term of Eq. (\ref{W}) describes the diffusion component of the
process, which draws the levels of neighboring agents closer. This
homogenizing force competes with the random idiosyncratic growth term 
$W_{i}(t)\eta _{i}(t)$. Random growth processes subject to diffusion such as
that of Eq. (\ref{W}) are well known in physics. In particular it is known 
\cite{DPRM} that the fluctuation properties of Eq. (\ref{W}) when $D$ is
larger than a critical value $D_{c}$ are qualitatively different to those
when $D<D_{c}$.

\begin{figure}[tbp]
\includegraphics[width=0.9\columnwidth]{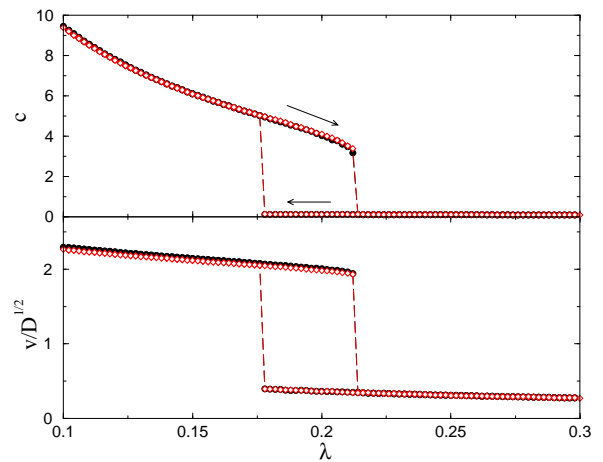}
\caption{Mean degree $c$ (top) and growth rate $v$ (bottom) as a function of 
$\lambda $ computed in numerical simulations of a discretized
version of the model with Eq. (\ref{W}). More precisely, we iterate
the equation $h_{i}(t+\Delta t)=\max_{j}h_{j}(t)+r_{i}$ where $j$ runs on
the local neighborhood of $i$, including $i$, $\Delta t$ is a small time
interval, and $r_{i}(t)$ is a Gaussian variable with mean $0$ and variance 
$\Delta t$. This equation describes the strong-noise limit of Eq. 
(\ref{W}) and it is obtained by setting $h_{i}=D^{-1/2}\log W_{i}$ when $D\gg 1$
(and $D\gg \lbrack \log (\protect\nu \Delta t)]^{2}$). Here we use $\Delta
t=0.2$, $\epsilon =0.01$, $\bar{d}=0.4\protect\sqrt{D}$ and $n=800$ (solid circles) and $1600$ (open diamonds).  }
\label{figh}
\end{figure}

Choosing $d_{ij}=|\log W_{i}-\log W_{j}|$ and updating both the links and $W$s at comparable timescales, we have performed extensive numerical simulations of the induced
network dynamics. Fig. \ref{figh} reports typical results for a simple
discretized version of Eq. (\ref{W}) with $D>D_{c}$ (see caption of Fig. \ref{figh}). 
As in the two previous models, we find a discontinuous transition
between a sparse and a dense network state, characterized by hysteresis
effects. When the network is sparse, diffusion is ineffective in
homogenizing growth. Hence the distance $d_{ij}$ is typically beyond the
threshold $\bar{d}$, thus slowing down the link formation process. On the
other hand, with a dense network, diffusion rapidly succeeds in narrowing
the gaps between the $W_{i}$s of different nodes, which in turn has a
positive effect on network formation. As before, the phase transition and
hysteresis is a result of the positive feedback that exists between the
dynamics of the $W_{i}$ and the adjustment of the network. In the stationary
state we find that $W(t)\equiv \left\langle W_{i}(t)\right\rangle $ grows
exponentially in time, i.e. $\log W_i(t)\simeq vt$.  Notably, the
growth process is much faster (i.e. $v$ is much higher) in the dense network
equilibrium than in the sparse one, as shown in the upper panel of Fig. \ref
{figh}.

Finally, we note that when diffusion is very strong compared to the
idiosyncratic shocks in Eq. (\ref{W}) -- i.e. $\nu \gg \sqrt{D}$ -- we
expect a much smaller distance $d_{ij}$ between agents in the same component compared to agents in different components.  Thus the model becomes similar to the first one in this limit, in 
the same way the second model did for $\beta \to \infty$.

\section{Conclusion}
In this paper we have proposed a general theoretical setup to
study the dynamics of a social network that is flexible enough to admit a
wide variety of particular specifications. We have studied three such
specifications, each illustrating a distinct way in which the network
dynamics may interplay with the adjustment of node attributes. In all these
cases, network evolution displays the three features (sharp transitions,
resilience, and equilibrium co-existence) that empirical research has found
to be common to many social-network phenomena. Our analysis indicates that
these features arise as a consequence of the cumulative self-reinforcing
effects induced by the interplay of two complementary considerations. On the
one hand, there is the subprocess by which agent similarity is enhanced
across linked (or close-by) agents. On the other hand, there is the fact
that the formation of new links is much easier between similar agents. When
such a feedback process is triggered, it provides a powerful mechanism that
effectively offsets the link decay induced by volatility.

The similarity-based forces driving the dynamics of the model are at work in
many socio-economic environments. Thus, even though fruitful economic
interaction often requires that the agents involved display some
\textquotedblleft complementary diversity\textquotedblright\ in certain
dimensions (e.g. buyers and sellers), a key prerequisite is also that agents
can coordinate in a number of other dimensions (e.g. technological standards
or trading conventions). Analogous considerations arise as well in the
evolution of many other social phenomena (e.g. the burst of social
pathologies discussed above) that, unlike what is claimed e.g. by Crane \cite
{Crane}, can hardly be understood as a process of epidemic contagion on a 
\emph{given} network. It is by now well understood \cite{Bailey, PV} that
such epidemic processes do \emph{not} match the phenomenology (a)-(c)
reported in empirical research. Our model suggests that a satisfactory
account of these phenomena must aim at integrating both the dynamics 
\emph{on} the network with that \emph{of} the network itself as part of a
genuinely co-evolutionary process.

\section{Appendix}

We characterize the long run behavior of the network in terms of the
stationary degree distribution $P(k)$, which is the fraction of agents with 
$k$ neighbors. This corresponds to approximating the network with a random
graph (see \cite{randomGraph}), an approximation which is rather accurate in
the cases we discuss here. We focus on the limit $n\rightarrow \infty $, for
which the analysis is simpler, but finite size corrections can be studied
within this same approach. The degree distribution satisfies a master
equation \cite{Gardiner}, which is specified in terms of the transition
rates $w(k\rightarrow k\pm 1)$ for the addition or removal of a link, for an
agent linked with $k$ neighbors. While $w(k\rightarrow k-1)=\lambda k$
always takes the same form, the transition rate $w(k\rightarrow k+1)$ for
the addition of a new link depends on the particular specification of the
distance $d_{ij}$. For the first model $w(k\rightarrow k+1)=\epsilon $ if
the two agents are in different components and $w(k\rightarrow k+1)=1$ if
they are in the same.  In the large $n$ limit the latter case only occurs with some probability if
the graph has a giant component $\mathcal{G}$ which contains a finite
fraction $\gamma$ of nodes. For random graphs (see Ref. \cite{randomGraph}
for details) the fraction of nodes in $\mathcal{G}$ is given by $\gamma
=1-\phi (u)$ where $\phi (s)=\sum_{k}P(k)s^{k}$ is the generating function
and $u$ is the probability that a link, followed in one direction, does not lead to the giant
component. The latter satisfies the equation $u=\phi ^{\prime }(u)/\phi
^{\prime }(1)$.  Hence $u^{k}$ is the probability an agent with $k$ neighbours has no 
links connecting him to the giant component, and hence is itself not part of the giant component.  Then the rate of
addition of links, in the first model, takes the form 
\begin{equation*}
w(k\rightarrow k+1)=2[\epsilon +(1-\epsilon )\gamma (1-u^{k})],
\end{equation*}
where the factor $2$ comes because each node can either initiate or
receive a new link. The stationary state condition of the master
equation leads to the following equation for $\phi (s)$ 
\begin{equation}
\lambda \phi ^{\prime }(s)=2[\epsilon +(1-\epsilon )\gamma ]\phi
(s)-2(1-\epsilon )\gamma \phi (us)  \label{phis}
\end{equation}
which can be solved numerically to the desired accuracy. Notice that Eq. 
(\ref{phis}) is a self-consistent problem, because the parameters $\gamma $
and $u$ depend on the solution $\phi (s)$. The solution of this equation is
summarized in Fig. \ref{fig_giant}. Either one or three solutions are found,
depending on the parameters. In the latter case, the intermediate solution
is unstable (dashed line in Fig. \ref{fig_giant}), and it separates the
basins of attraction of the two stable solutions within the present mean
field theory. Numerical simulations reveal that the the mean field approach
is very accurate away from the phase transition although it overestimates the size of
the coexistence region.

\begin{figure}[tbp]
\includegraphics[width=0.9\columnwidth]{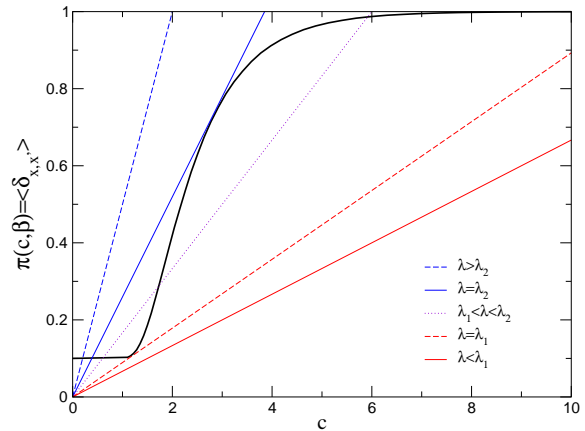}
\caption{Graphical solution for the stationary state of the coordination
model for $q=10$ and $\protect\beta=8$.}
\label{figpic}
\end{figure}

Now we turn to the second model, where each node displays one out of a finite
set of attributes. In order to simplify the analysis, we approximate the
prevailing network $g$ with a random graph with Poisson degree distribution
and average degree $c$, i.e. a graph where any given link $ij$ is present with
probability $c/(n-1)$. Though not exact, this approximation is rather accurate
as confirmed by numerical simulations, and it allows us to clarify the
behavior of the model in a simple and intuitive way. (A more precise
solution, which relies on a more accurate description of the network
topology can also be derived, yielding no essential differences.) The
solution of the Potts model on random graphs of Ref. \cite{DGM,EM} (with temperature $T=1/(2 k_B \beta)$) allows us
to compute the probability that two randomly chosen nodes $i$ and $j$ have $
x_{i}=x_{j}$. Given the Poisson approximation, such a probability is given
by a function $\pi (c,\beta )=\langle {\delta _{x_{i},x_{j}}}\rangle $ of
the average degree $c$ and $\beta $, as plotted in Fig. \ref{figpic}.  
Equalizing the link destruction and formation rate $\lambda c/2 = \pi (c,\beta )$
yields an equation for the equilibrium values of $c$, for any given $\beta$
. A graphical approach shows that when $\lambda >\lambda _{2}$ there is a
single solution, representing a sparse network. At $\lambda _{2}$ two other 
solutions arise, one of which is unstable as above. At a further point $
\lambda _{1}$ the sparse-network solution merges with the unstable one
and both disappear for $\lambda <\lambda _{1}$, leaving only a solution with
a stable and dense network. This reproduces the same phenomenology observed
in the numerical simulations of the second model, which is also
qualitatively similar to that presented in Fig. \ref{fig_giant} for the
first model.

\end{document}